\documentclass[journal,12pt,twocolumn]{IEEEtran}
\usepackage{graphicx}
\usepackage{diagbox}
\usepackage{color} 
\usepackage{caption}
\usepackage{subcaption}
\usepackage{etoolbox}
\usepackage{amssymb,amsmath,amsfonts,amsthm}
\usepackage{tikz}
\usepackage{multirow}
\usepackage[affil-it]{authblk}
\usepackage{url}
\usepackage{hyperref}

\theoremstyle{plain}

\begin{document}

\title{WiBall: A Time-Reversal Focusing Ball Method for Indoor Tracking}

\author[${\star \dagger}$]{Feng Zhang}
\author[${\star \dagger}$]{Chen Chen}
\author[${\star \dagger}$]{Beibei Wang}
\author[${\dagger}$]{Yi Han}
\author[${\star \dagger}$]{K. J. Ray Liu}

\affil[${\star}$]{University of Maryland, College Park, MD, 20742, USA.}
\affil[${\dagger}$]{Origin Wireless, Inc., Suite 1070, 7500 Greenway Center Drive, MD, 20770, USA.}
\affil[$\star$]{Email: \{fzhang15, cc8834, bebewang, kjrliu\}@umd.edu}
\affil[$\dagger$]{Email: yi.han@originwireless.net}

\maketitle

\begin{abstract}

With the development of the Internet of Things technology, indoor tracking has become a popular application nowadays, but most existing solutions can only work in line-of-sight scenarios, or require regular re-calibration. In this paper, we propose WiBall, an accurate and calibration-free indoor tracking system that can work well in non-line-of-sight based on radio signals. WiBall leverages a stationary and location-independent property of the time-reversal focusing effect of radio signals for highly accurate moving distance estimation. Together with the direction estimation based on inertial measurement unit and location correction using the constraints from the floorplan, WiBall is shown to be able to track a moving object with decimeter-level accuracy in different environments. Since WiBall can accommodate a large number of users with only a single pair of devices, it is low-cost and easily scalable, and can be a promising candidate for future indoor tracking applications.

\end{abstract}


\section{Introduction}
\label{sec:Introduction}
With the proliferation of the Internet of Things (IoT) applications, Indoor Positioning and Indoor Navigation (IPIN) has received an increasing attention in recent years. Technavio forecasts the global IPIN market to grow to USD 7.8 billion by 2021~\cite{report17market}, and more than ever before, enterprises of all sizes are investing in IPIN technology to support a growing list of applications, including patient tracking in hospitals, asset management for large groceries, workflow automation in large factories, navigation in malls, appliance control, etc.

Although Global Positioning System (GPS) can achieve good accuracy with a low cost in outdoor real-time tracking, such a good balance between the cost and performance has not been realized for indoor tracking yet~\cite{chen2017achieving}. Existing solutions can be classified into four categories: vision based, audio based, radio based, and inertial measurement unit (IMU) based. Vision-based approaches, such as camera~\cite{werner2011indoor}, laser~\cite{mautz2011survey}, infrared~\cite{gorostiza2011infrared}, etc., suffer from high cost of deployment and equipment, sophisticated calibrations and limited coverage, although a very high accuracy can be achieved. The acoustic-based schemes~\cite{rishabh2012indoor} can only cover a limited range and are not scalable to a large number of users. The performance of radio-based approaches, such as RADAR~\cite{barton2004radar} and UWB localization systems~\cite{kuhn2008high} \cite{lee2002ranging}, is severely affected by the non-line-of-sight (NLOS) multipath propagation which is unavoidable for a typical indoor environment. The localization accuracy of IMU-based methods~\cite{wang2012no}\cite{yang2015mobility} is mainly limited by the poor estimation of moving distance and the drifting of the gyroscope.

Due to the wide deployment of WiFi in indoors, various indoor localization systems based on WiFi have been proposed, as summarized in Table~\ref{tab:Brief_Summary_Existing_WiFi_Approaches}. In general, these works can be classified into two categories: modeling-based approach and fingerprinting-based approach. The features utilized in these approaches can be obtained either from the MAC layer information, e.g., receive signal strength indicator (RSSI) readings and the timestamps of the received packets at the receiver (RX), or from the PHY information, e.g., the channel state information (CSI).

In the modeling-based schemes, either the distance~\cite{giustiniano2011caesar}\cite{sen2015bringing}\cite{bahl2000radar} or the angle~\cite{kotaru2015spotfi}\cite{gjengset2014phaser}\cite{sen2013avoiding} between an anchor point and the device can be estimated and the device can be localized by performing geometrical triangulation. The distance between the anchor point and the device can be estimated from the decay of RSSI~\cite{youssef2005horus} or from the time of arrival (ToA) of the transmitted packets which can be extracted from the timestamps of the received packets~\cite{xiong2015tonetrack}. The angle in between can be obtained by examining the features of the CSI received by multiple receive antennas, and then, the angle of arrival (AoA) of the direct path to the target can be calculated. ToA-based methods typically require synchronization between the anchor point and the device and thus are very sensitive to timing offsets~\cite{golden2007sensor}; AoA-based methods require an array of phased antennas which are not readily available in commercial WiFi chips~\cite{gjengset2014phaser}. The main challenges for the modeling-based approaches are the blockage and reflection of the transmitted signal since only the signal coming from the direct path between the anchor point and the device is useful for localization.

The fingerprinting-based schemes consist of an offline phase and an online phase. During the offline phase, features associated with different locations are extracted from the WiFi signals and stored in a database; in the online phase, the same features are extracted from the instantaneous WiFi signals and compared with the stored features so as to classify the locations. The features can be obtained either from the vector of RSSIs~\cite{youssef2005horus}\cite{castro2001probabilistic} or the detailed CSI~\cite{wang2015deepfi}\cite{wang2015phasefi}\cite{wu2015time} from a specific location to all the anchor points in range. A major drawback of the fingerprinting-based approaches lies in that the features they use are susceptible to the dynamics of the environment. For example, the change of furniture or the status of doors may have a severe impact on these features and the database of the mapped fingerprints need to be updated before it can be used again. In addition, the computational complexity of the fingerprinting-based approaches scales with the size of the database and thus they are not feasible for low-latency applications, especially when the number of the collected fingerprints is large.

\begin{table}[tb]
\scriptsize
\centering
\caption{A brief summary of the WiFi-based approaches}
\label{tab:Brief_Summary_Existing_WiFi_Approaches}
\begin{tabular}{|l|l|l|}
\hline                & Method & Existing Solutions                 \\ \hline
\multirow{4}{*}{\begin{tabular}[c]{@{}l@{}}Modeling-\\ based\end{tabular}}          & ToA    & CAESAR~\cite{giustiniano2011caesar}, ToneTrack~\cite{xiong2015tonetrack} \\ \cline{2-3}
                                                                                & AoA    & ArrayTrack~\cite{xiong2013arraytrack}, SpotFi~\cite{kotaru2015spotfi}, Phaser~\cite{gjengset2014phaser} \\ \cline{2-3}
                                                                                & RSSI    & RADAR~\cite{bahl2000radar}                     \\ \cline{2-3}
                                                                                & CSI    & FILA~\cite{wu2013csi}                       \\ \hline
\multirow{2}{*}{\begin{tabular}[c]{@{}l@{}}Fingerprinting-\\ based \end{tabular}} & RSSI    & Horus~\cite{youssef2005horus}, Nibble~\cite{castro2001probabilistic}          \\ \cline{2-3}
                                                                                & CSI    & PinLoc~\cite{sen2011precise}, TRIPS~\cite{wu2015time}, DeepFi~\cite{wang2015deepfi}   \\ \hline
\end{tabular}
\end{table}

In sum, the performance of most existing solutions for indoor localization degrades dramatically under NLOS conditions, which are common usage scenarios though. Even with a significant overhead in the manual construction of the database, the fingerprinting-based approaches still fail to achieve a decimeter-level accuracy. Therefore, indoor location-based services are not provided as widespread as expected nowadays, which motivates us to design a highly accurate and robust indoor tracking system even without requiring specialized infrastructure.

In this paper, we propose WiBall for indoor tracking that can work well in both NLOS and line-of-sight (LOS) scenarios and is robust to environmental dynamics as well. WiBall estimates the incremental displacement of the device at every moment, and thus, it can track the trace of the device in real time. WiBall adopts a completely new paradigm in the moving distance estimation, which is built on the proposed discovered physical phenomenon of radio signals. In the past, the moving distance estimation can be done by analyzing the output of IMU that is attached to the moving object. Accelerometer readings are used to detect walking steps and then, the walking distance can be estimated by multiplying the number of steps with the stride length~\cite{wang2012no}. However, pedestrians often have different stride lengths that may vary up to $40\%$ even at the same speed, and $50\%$ with various speeds of the same person~\cite{yang2015mobility}. Thus calibration is required to obtain the average stride lengths for different individuals, which is impractical in real applications and thus has not been widely adopted. The moving distance can also be estimated by analyzing radio signals that are affected by the movement of the device. Various methods have been proposed based on the estimation of the maximum Doppler frequency, such as level crossing rate methods~\cite{park2003level}, covariance based methods~\cite{sampath1993estimation}\cite{xiao2001mobile}, and wavelet-based methods~\cite{narasimhan1999speed}. However, the performance of these estimators is unsatisfactory in practical scenarios. For example, the approach in \cite{xiao2001mobile} can only differentiate whether a mobile station moves with a fast speed ($\geq30$km/h) or with a slow speed ($\leq5$km/h).

In WiBall, a new scheme for moving distance estimation based on the time-reversal (TR) resonating effect~\cite{lerosey2004time}\cite{wang2011green} is proposed. TR is a fundamental physical resonance phenomenon that allows people to focus the energy of a transmitted signal at an intended focal spot, both in the time and spatial domains. The research of TR can be traced back to the 1950s when it was first utilized to align the phase differences caused by multipath fading during long-distance information transmissions. The TR resonating effect was first observed in a practical underwater propagation environment~\cite{roux1997time} that the energy of a transmitted signal can be refocused at the intended location because by means of TR the RX recollects multipath copies of a transmitted signal in a coherent matter.

In this paper, we present a new discovery that the energy distribution of the TR focusing effect exhibits a location-independent property, which is only related to the physical parameters of the transmitted EM waves. This is because the number of multipath components (MPC) in indoors is so large that the randomness of the received energy at different locations can be averaged out as a result of the law of large numbers. Based on this location-independent feature, WiBall can estimate the moving distance of the device in a complex indoor environment without requiring any pre-calibration procedures. To cope with the cumulative errors in distance estimation, WiBall incorporates the constraints imposed by the floorplan of buildings and corrects the cumulative errors whenever a landmark, such as a corner, hallway, door, etc., is met. Combining the improved distance estimator and the map-based error corrector, the proposed WiBall is shown to be able to achieve decimeter-level accuracy in real-time tracking regardless of the moving speed and environment.

The rest of the paper is organized as follows. Section \ref{sec:TR_principle_for_distance_estimation} introduces the system architecture of WiBall followed by a discussion on the TR principle for distance estimation. Section \ref{sec:Direction_Map} presents an IMU-based moving direction estimator and a map-based localization corrector, which can correct the accumulated localization error.  Experimental evaluation is shown in Section \ref{sec:Experiment_Evaluations} and Section \ref{sec:Concluding_Remarks} concludes the paper.

\section{TR Focusing Ball Method for Distance Estimation}
\label{sec:TR_principle_for_distance_estimation}
In this section, we first introduce the overall system architecture of WiBall and the TR radio system. Then, we derive the analytical energy distribution of the TR focal spot. We show that the energy distribution is location-independent and can be used to estimate distance. Last, we discuss the TR-based distance estimator.

\subsection{Overview of WiBall}
WiBall consists of a transmitter (TX) broadcasting beacon signals periodically to all the RXs being tracked. WiBall estimates the paths that the RX travels, i.e., the location of the RX $\vec{\mathbf{x}}$ at time $t_i$ is estimated as
\begin{eqnarray}
\vec{\mathbf{x}}(t_i) = \vec{\mathbf{x}}(t_{i-1}) + \vec{\Delta}(t_{i}),
\end{eqnarray}
where $\vec{x}(t_{i-1})$ represents the location of the RX at the previous time $t_{i-1}$, and $\Delta(t_i)$ is the incremental displacement. The magnitude of $\vec{\Delta}(t_{i})$, denoted as $d(t_i)$, and the angle of $\vec{\Delta}(t_{i})$, denoted as $\theta(t_i)$, correspond to the moving distance and the change of moving direction of the RX from $t_{i-1}$ to $t_{i}$, respectively.

WiBall estimates the moving distance $d(t_i)$ based on the TR resonating effect, which can be obtained from the CSI measurements at the RX. The estimation of $\theta(t_i)$ is based on the angular velocity and gravity direction from IMU, which is a built-in module for most smartphones nowadays.

\subsection{TR Radio System}
Consider a rich-scattering environment, e.g., an indoor or metropolitan area, and a wireless transceiver pair each equipped with a single omnidirectional antenna. Given a large enough bandwidth, the MPCs in a rich-scattering environment can be resolved into multiple taps in discrete-time and let $h(k;\vec{T}\rightarrow \vec{R}_0)$ denote the $k$-th tap of the channel impulse response (CIR) from $\vec{T}$ to $\vec{R}_0$, where $\vec{T}$ and $\vec{R}_0$ denotes the coordinates of the TX and RX, respectively. In the TR transmission scheme, the RX at $\vec{R}_0$ first transmits an impulse and the TX at $\vec{T}$ captures the CIR from $\vec{R}_0$ to $\vec{T}$. Then the RX at $\vec{T}$ simply transmits back the time-reversed and conjugated version of the captured CIR, i.e., $h^*(-k;\vec{R}_0\rightarrow \vec{T})$, where $*$ denotes complex conjugation. With channel reciprocity, i.e., the forward and backward channels are identical~\cite{wu2015time}, the received signal at any location $\vec{R}$ when the TR waveform $h^*(-k;\vec{R}_0\rightarrow \vec{T})$ is transmitted can be written as~\cite{el2010experimental}
\begin{equation}
\label{equ: receive_signal}
s(k;\vec{R})=\sum\limits_{l=0}\limits^{L-1}h(l;\vec{T}\rightarrow\vec{R})h^*(l-k;\vec{R}_0\rightarrow\vec{T}),
\end{equation}
where $L$ is the number of resolved multipaths in the environment. When $\vec{R}=\vec{R}_0$ and $k=0$, we have $s(k;\vec{R})=\sum_{l=0}^{L-1}|h(l,\vec{T}\rightarrow\vec{R}_0)|^{2}$ with all MPCs added up coherently, i.e., the signal energy is refocused at a particular spatial location at a specific time instance. This phenomenon is termed as the TR spatial-temporal resonating effect~\cite{lerosey2004time}.

To study the TR resonating effect in the spatial domain, we fix time index $k=0$ and define the TR resonating strength (TRRS) between the CIRs of two locations $\vec{R}_0$ and $\vec{R}$ as the normalized energy of the received signal when the TR waveform for location $\vec{R}_0$ is transmitted:
\begin{eqnarray}
\label{equ: TR resonating strength}
& & \eta(\mathbf{h}(\vec{R}_0),\mathbf{h}(\vec{R})) \nonumber\\
\!\!\!\!&=&\!\!\!\! \left| \frac{s(0;\vec{R})}{\sqrt{\sum\limits_{l_1=0}\limits^{L-1} |h(l_1;\vec{T}\rightarrow\vec{R}_0)|^2}\sqrt{\sum\limits_{l_2=0}\limits^{L-1} |h(l_2;\vec{T}\rightarrow\vec{R})|^2}} \right|^2,
\end{eqnarray}
where we use $\mathbf{h}(\vec{R})$ as an abbreviation of $h(l;\vec{T}\rightarrow\vec{R}),\;l=0,...,L-1$, when $\vec{T}$ is fixed. Note that the range of TRRS is normalized to be $[0,1]$ and TRRS is symmetric, i.e., $\eta(\mathbf{h}(\vec{R}_0),\mathbf{h}(\vec{R}))=\eta(\mathbf{h}(\vec{R}),\mathbf{h}(\vec{R}_0))$.

\begin{figure}[t]
    \centering
    \begin{subfigure}[b]{0.45\textwidth}
        \includegraphics[width=\textwidth]{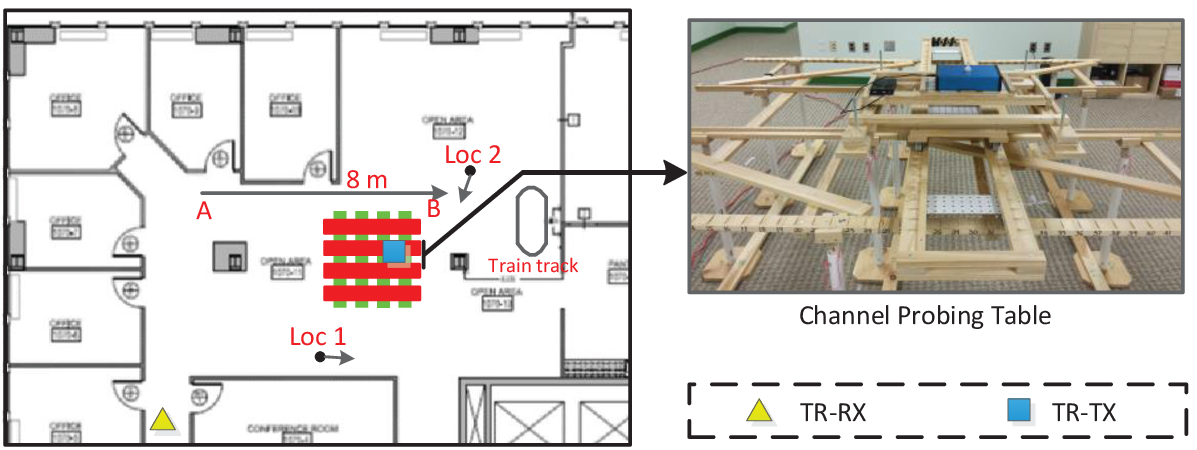}
        \caption{\label{fig:TR_prototype} TR prototype and channel probing platform.}
    \end{subfigure}
    ~
    \begin{subfigure}[b]{0.23\textwidth}
        \includegraphics[width=\textwidth]{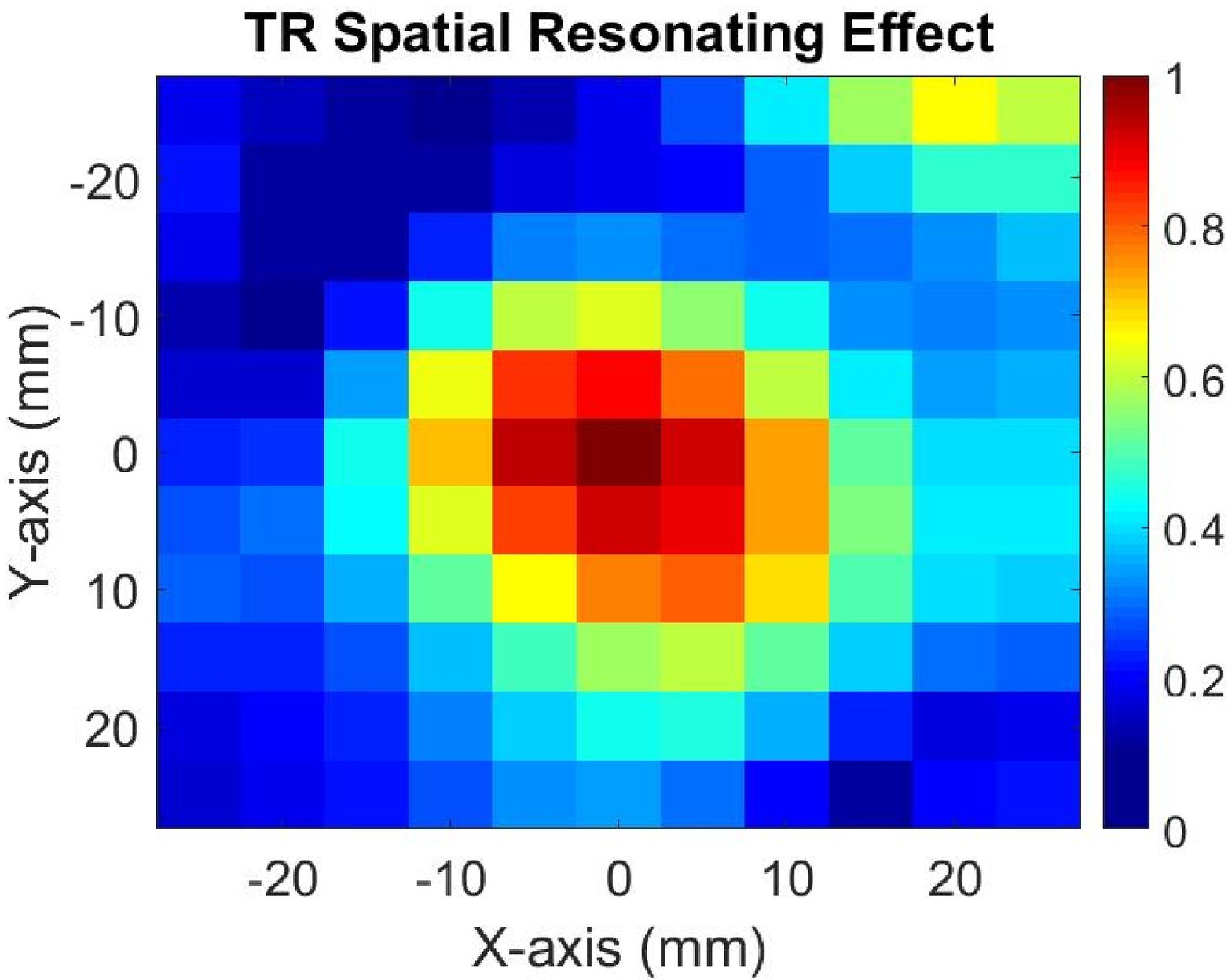}
        \caption{\label{fig:Spatial_Resonating_Effect} TRRS in spatial domain}
    \end{subfigure}
    ~
    \begin{subfigure}[b]{0.23\textwidth}
        \includegraphics[width=\textwidth]{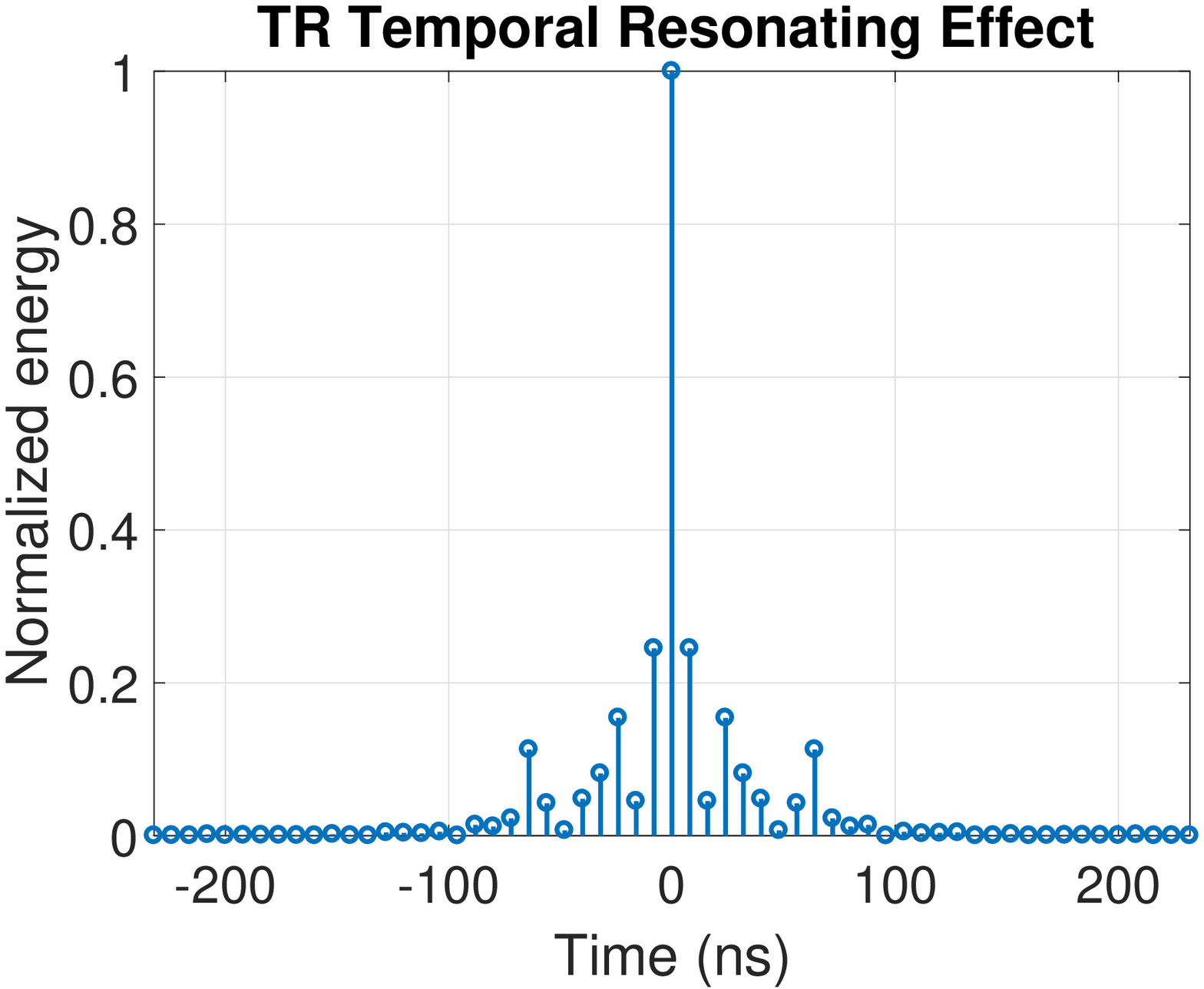}
        \caption{\label{fig:Temporal_Resonating_Effect} TRRS in time domain}
    \end{subfigure}
    \caption{TR prototype and the environment of the measurement, the TRRS distribution in the spatial domain, and the normalized energy of the received signal at the focal spot $\vec{R}_0$ in the time domain.}
\end{figure}

We built a pair of customized TR devices to measure the TRRS at different locations, as shown in Fig.~\ref{fig:TR_prototype}. The devices operate at $f_0=5.8$GHz ISM band with $125$MHz bandwidth, and the corresponding wavelength is $\lambda=c/f_0=5.17$cm. The RX is placed on a $5\;\textrm{cm}\times 5\;\textrm{cm}$ square area above a channel probing table with $0.5$cm resolution, and the center of the square is set to be the focal spot $\vec{R}_0$. The TRRS distribution around $\vec{R}_0$ in the spatial domain and the normalized received energy at $\vec{R}_0$ in the time domain are shown in Fig.~\ref{fig:Spatial_Resonating_Effect} and Fig.~\ref{fig:Temporal_Resonating_Effect}, respectively. As we can see from the results, the received energy is concentrated around $\vec{R}_0$ both in spatial and time domains almost symmetrically, which shows that a bandwidth of $125$MHz is able to achieve the TR resonating effect in a typical indoor environment.

\subsection{Energy Distribution of TR Focal Spot}
\begin{figure}[htbp]
\centering
\includegraphics[width=0.4\textwidth]{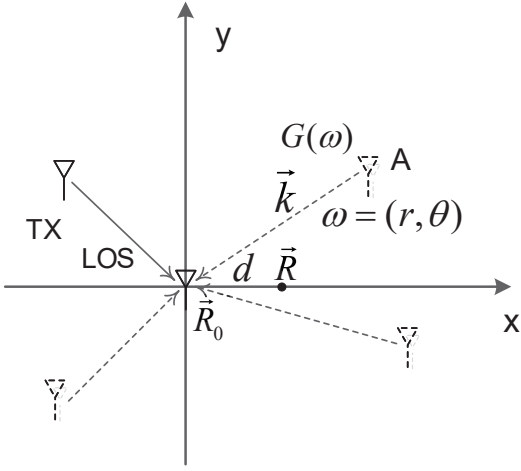}
\caption{Illustration of the polar coordinates in the analysis.}
\label{fig:Bessel_Explanation}
\end{figure}

Assume that all the EM waves propagate in a far-field zone, and then each MPC can be approximated by a plane EM wave. For the purpose of illustration, each MPC can be represented by a point, e.g., point $A$, as shown in Fig.~\ref{fig:Bessel_Explanation}, where $r$ stands for the total traveled distance of the multipath, $\theta$ denotes the direction of arrival of the multipath, and $G(\omega)$ denotes the power gain with $\omega=(r,\theta)$. In a rich-scattering environment, we can also assume that $\omega$ is uniformly distributed in the space and the total number of MPCs is large. When a vertically polarized antenna is used, only the EM waves with the direction of electric field orthogonal to the horizontal plane are collected. Then, the received signal is just a scalar sum of the electric field of the impinging EM waves along the vertical direction. In the sequel, without loss of generality, we only consider the TRRS distribution in the horizontal plane since its distribution in the vertical plane is similar.

For a system with bandwidth $B$, two MPCs would be divided into different taps of the measured CIR as long as the difference of their time of arrival is larger than the sampling period $1/B$, that is, any two MPCs with a difference of their traveled distances larger than $c/B$ can be separated. With a sufficiently large system bandwidth $B$, the distance resolution $c/B$ of the system is so small that all of the MPCs with significant energy can be separated in the spatial domain, i.e., each significant MPC can be represented by a single tap of a measured CIR. Assume that the distribution of the energy of each MPC is uniform in direction $\theta$, i.e., the distribution of $G(\omega)$ is only a function of $r$. Then the energy of MPCs coming from different directions would be approximately the same when the number of MPCs is large. Mathematically, for any point $\vec{R}$ in a source-free region with constant mean electric and magnetic fields, the channel impulse response when a delta-like pulse is transmitted can be written as~\cite{el2010experimental}
\begin{eqnarray}
\label{equ: detailed cir}
\!\!\!\!& & h(t;\vec{T}\rightarrow \vec{R}) \nonumber \\
\!\!\!\!\!\!\!\!\!&=&\!\!\!\!\! \sum_{\omega\in\Omega}\!\!G(\omega)q(t\!-\!\tau(\omega))e^{i(2\pi f_0(t-\tau(\omega))-\phi(\omega)-\vec{k}(\omega)\cdot\vec{R})},
\end{eqnarray}
where $q(t)$ is the pulse shaper, $\tau(\omega)=r/c$ is the propagation delay of the MPC $\omega$, $f_0$ is the carrier frequency, $\Omega$ is the set of MPCs, $\phi(\omega)$ is the change of phase due to reflections and $\vec{k}(\omega)$ is the wave vector with amplitude $k=c/f_{0}$. Accordingly, the $l$-th tap of a sampled CIR at location $\vec{R}$ can be expressed as
\begin{eqnarray}
\!\!\!\!& & h(l;\vec{T}\rightarrow\vec{R})\nonumber\\
\!\!\!\!\!\!\!&=&\!\!\!\!\!\!\!\!\!\!\!\!\!\!\!\!\! \sum_{\tau(\omega)\in[lT-\frac{T}{2},lT+\frac{T}{2})}\!\!\!\!\!\!\!\!\!\!\!\!\!\!\!\!G(\omega)q(\Delta\tau(l,\omega))e^{i(2\pi f_0\Delta\tau(l,\omega)-\phi(\omega)-\vec{k}(\omega)\cdot\vec{R})}
\end{eqnarray}
where $T$ is the channel measurement interval and $\Delta\tau(l,\omega)=lT-\tau(\omega)$ for $l=0,1,...,L-1$. When the TR waveform $h^*(-l;\vec{R}_0\rightarrow\vec{T})$ is transmitted, the corresponding received signal at the focal spot $\vec{R}_0$ can be written as
\begin{eqnarray}
\label{eq:pre TRRS}
& &s(0;\vec{R}) \nonumber\\
\!\!\!\!\!\!\!\!\!\!\!\!\!&=&\!\!\!\!\! \sum_{l=1}^L \left| \sum_{\tau\in[lT-\frac{T}{2},lT+\frac{T}{2})}\!\!\!\!\!\!\!\!\!\!\!\!\!\! G(\omega)q(\Delta\tau(l\!,\!\omega))e^{i(2\pi f_0\Delta\tau(l\!,\omega)\!-\!\phi(\omega))}\! \right|^2.
\end{eqnarray}
\eqref{eq:pre TRRS} shows that the MPCs with propagation delays $\tau(\omega)\in[lT-\frac{T}{2},lT+\frac{T}{2})$ for each $l$ would be merged into one single tap, and the signals coming from different taps would add up coherently while the MPCs within each sampling period $T$ would add up incoherently. It indicates that the larger the bandwidth, the larger the TR focusing gain that can be achieved, since more MPCs can be aligned and added up coherently. When the bandwidth is sufficiently large, the received signal at each point $\vec{R}$ can be approximated as
\begin{eqnarray}
s(0;\vec{R}) \approx \sum_{l=1}^L |G(\omega)q(\Delta\tau(l,\omega))|^2 e^{-i\vec{k}(\omega)\cdot(\vec{R}-\vec{R}_0)}.
\end{eqnarray}
When a rectangular pulse shaper is used, i.e., $q(t)=1$ for $t\in[-\frac{T}{2},\frac{T}{2})$ and $q(t)=0$ otherwise, under the above symmetric scattering assumption the received signal $s(0;\vec{R})$ can thus be approximated as
\begin{eqnarray}
\label{equ: Bessel_Approximation}
s(0;\vec{R})&=&\sum\limits_{\omega\in\Omega}|G(\omega)|^2e^{-i\vec{k}\cdot(\vec{R}-\vec{R}_0)} \nonumber\\
&\approx& \int_0^{2\pi} P(\theta)e^{-ikd\cos(\theta)}d\theta \nonumber\\
&=& PJ_0(kd),
\end{eqnarray}
where the coordinate system in Fig.~\ref{fig:Bessel_Explanation} is used, $\Omega$ stands for the set of all significant MPCs, $J_0(x)$ is the $0^{th}$-order Bessel function of the first kind, and $d$ is the Euclidean distance between $\vec{R}_0$ and $\vec{R}$. Here we use a continuous integral to approximate the discrete sum and $P(\theta)=P$ denotes the density of the energy of MPCs coming from direction $\theta$. For $\vec{R}=\vec{R}_0$, it degenerates to the case of $d=0$ and thus $s(0;\vec{R}_0)\approx P$. Since the denominator of \eqref{equ: TR resonating strength} is the product of the energy received at two focal spots, it would converge to $P^2$. At the same time, the numerator is approximately $P^2J_0^2(kd)$ as discussed above. As a result, the TRRS defined in \eqref{equ: TR resonating strength} can be approximated as
\begin{equation}
\label{equ:besselfunction}
\eta(\mathbf{h}(\vec{R}_0),\mathbf{h}(\vec{R})) \approx J_0^2(kd).
\end{equation}
In the following, since the theoretic approximation of the TRRS distribution only depends on the distance between two points, we use $\bar{\eta}(d)=J_0^2(kd)$ to stand for the approximation of TRRS between two points with distance $d$.

\begin{figure}
    \centering
    \begin{subfigure}[b]{0.45\textwidth}
        \includegraphics[width=\textwidth]{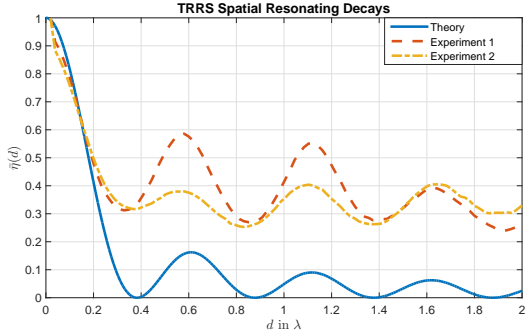}
        \caption{\label{fig:Resonating_Theory_Experiment} Comparison of the TRRS distribution between experimental results and the theoretical result. Experiments 1 and 2 correspond to Location 1 and Location 2 respectively in Fig.~\ref{fig:TR_prototype}.}
    \end{subfigure}
    ~
    \begin{subfigure}[b]{0.41\textwidth}
        \includegraphics[width=\textwidth]{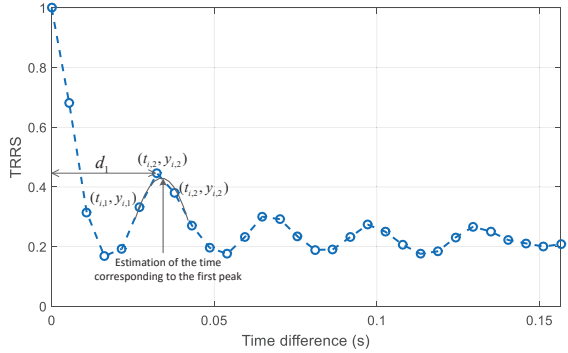}
        \caption{\label{fig:Speed_Estimation_Algorithm}Illustration of the proposed TR-based speed estimation algorithm.}
    \end{subfigure}
    \caption{The distributions of TRRS.}
\end{figure}

To evaluate the above theoretic approximation, we also built a mobile channel probing platform equipped with stepping motors that can control the granularity of the CIR measurements precisely along any predefined direction. Extensive measurements of CIRs from different locations have been collected in the environment shown in Fig.~\ref{fig:TR_prototype}. Fig.~\ref{fig:Resonating_Theory_Experiment} shows two typical experimental results measured at Location 1 and Location 2 with a separation of $10$m as shown in Fig.~\ref{fig:TR_prototype}. The distance $d$ away from each predefined focal spot increases from $0$ to $2\lambda$ with a resolution of $1$mm. The measured TRRS distribution functions agree with the theoretic approximation quite well in the way that the positions of the peaks and valleys in the measured curves are almost the same as those of the theoretic curve. Although Locations 1 and 2 are far apart, the measured TRRS distribution functions exhibit similar damping pattern when the distance $d$ increases.

We also observe that the measured TRRS distribution functions are far above $0$. This is due to the contribution of the direct path between the TR devices. Therefore, the energy density function $P(\theta)$ in \eqref{equ: Bessel_Approximation} consists of a term which is symmetric in direction due to NLOS components and another term which is asymmetric in direction due to LOS components. As a result, the TRRS is indeed a superposition of $J_0^2(kd)$ and some unknown function which is the result of the asymmetric energy distribution of MPCs in certain directions. Since the pattern of $J_0^2(kd)$, embedded in the TRRS distribution function, is location-independent, we can exploit this feature for speed estimation.

\begin{figure}[htbp]
    \centering
    \begin{subfigure}[b]{0.23\textwidth}
        \includegraphics[width=\textwidth]{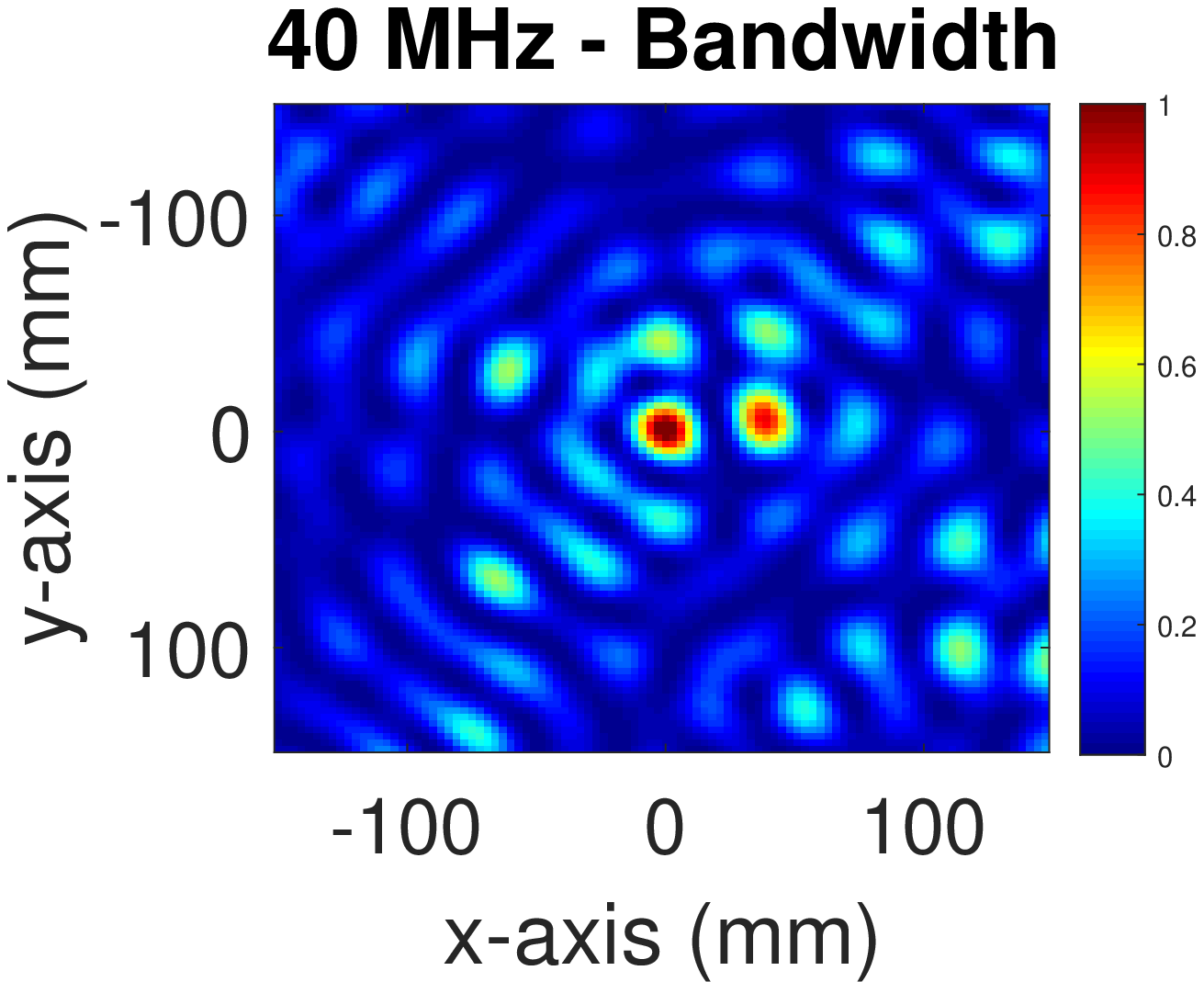}
        \caption{\label{fig:sim_40}40 MHz bandwidth.}

    \end{subfigure}
    ~
    \begin{subfigure}[b]{0.23\textwidth}
        \includegraphics[width=\textwidth]{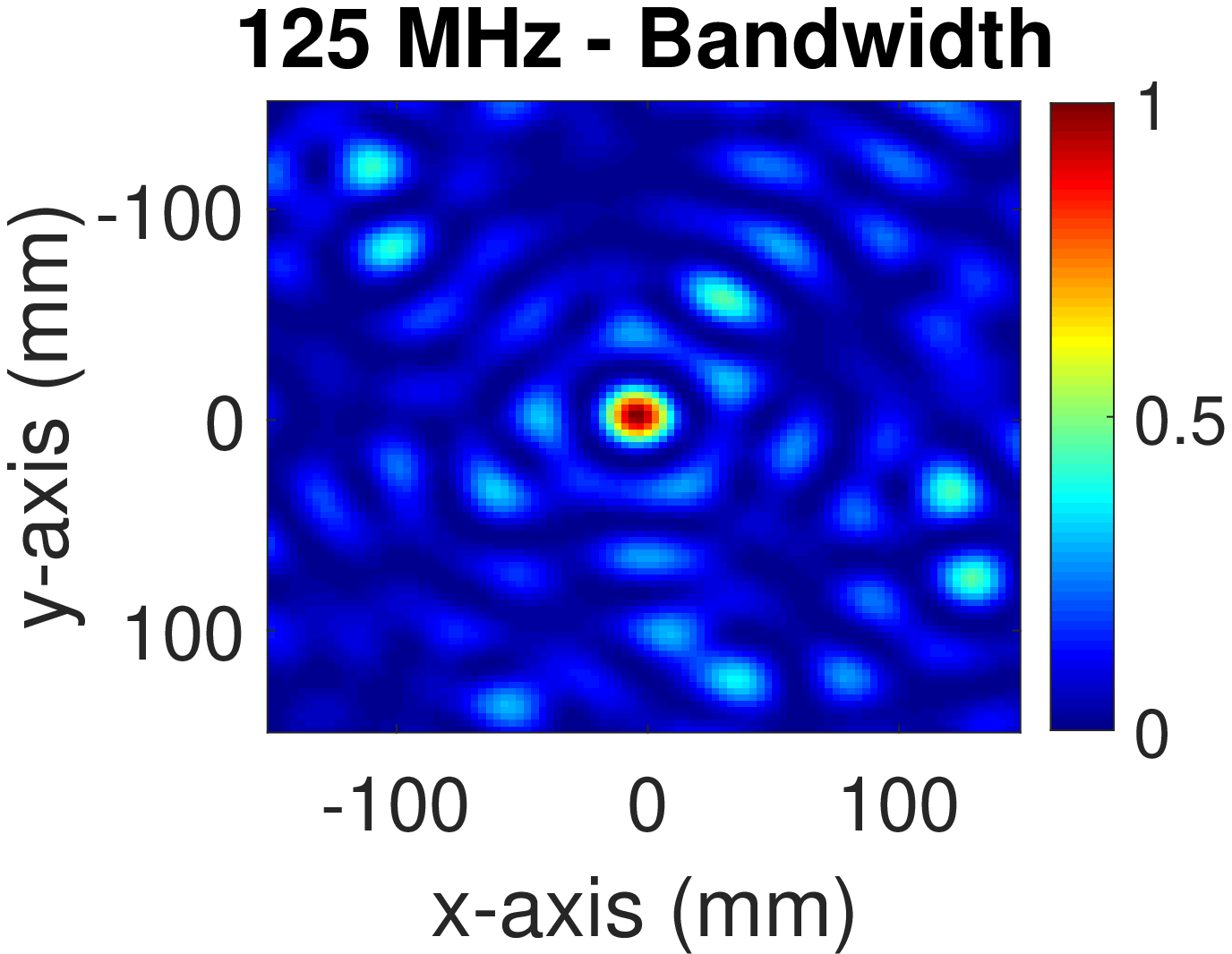}
        \caption{\label{fig:sim_125} 125 MHz bandwidth.}

    \end{subfigure}
    ~
    \begin{subfigure}[b]{0.23\textwidth}
        \includegraphics[width=\textwidth]{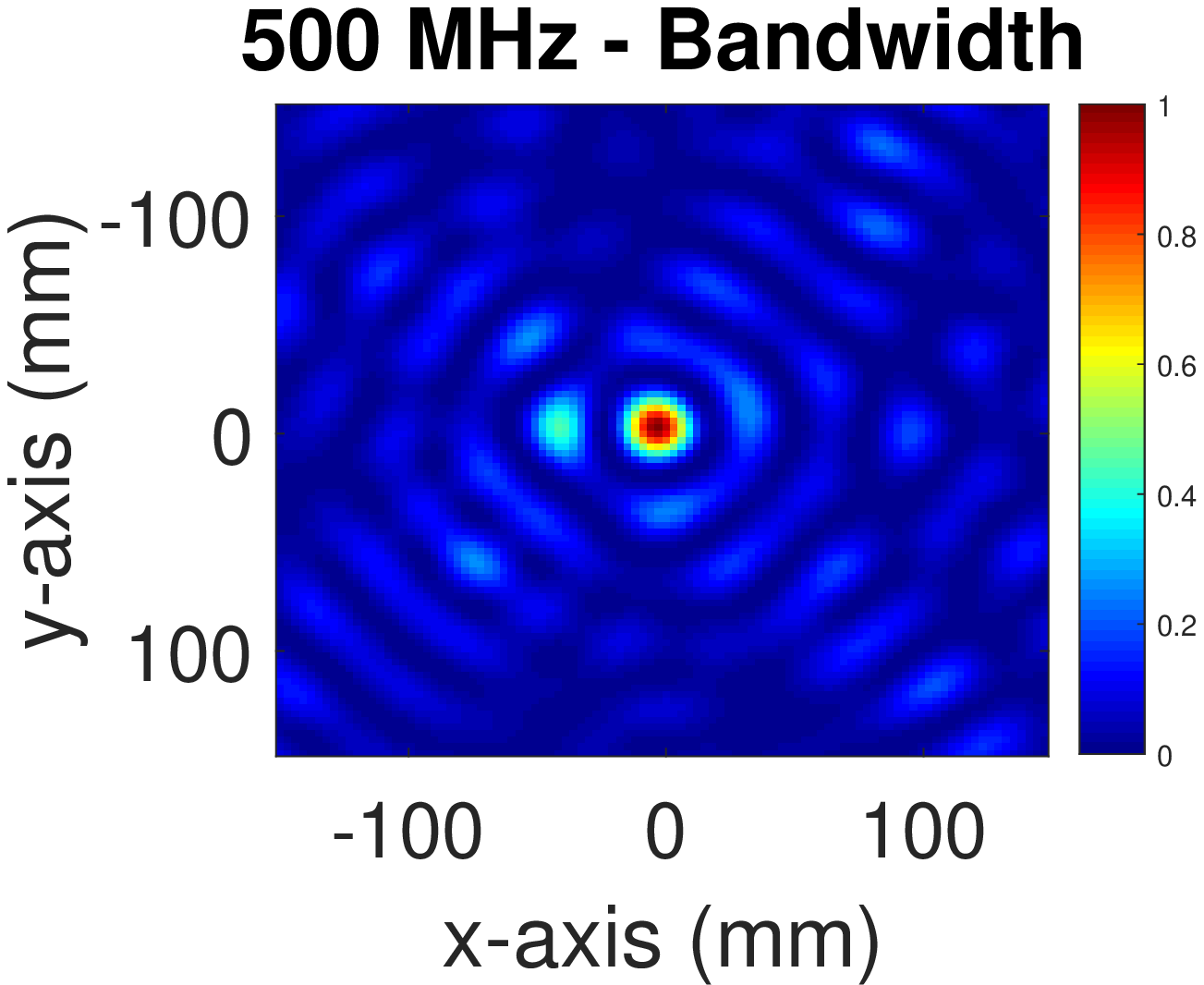}
\caption{\label{fig:sim_500} 500 MHz bandwidth.}
    \end{subfigure}
    \caption{\label{fig:Numerical_Simulations} Numerical simulations of the distributions of TRRS with varying bandwidth.}
\end{figure}

A numerical simulation using a ray-tracing approach is also implemented to study the impact of bandwidth on TRRS distribution. In the simulation, the carrier frequency of the transmitted signals is set to be $5.8$ GHz. $200$ scatterers are uniformly distributed in a $7.5$ m by $7.5$ m square area. The reflection coefficient is distributed uniformly and independently in $(0,1)$ for each scatterer. The distance between the TX and RX is $30$ m and the RX (focal spot) is set to be the center of the square area. Fig.~\ref{fig:Numerical_Simulations} shows the distributions of TRRS around the focal spot when the system bandwidth $40$ MHz, $125$ MHz and $500$ MHz, respectively. As we can see from the results, as the bandwidth increases, the distribution of TRRS in the horizontal plane becomes more deterministic-like and converges to the theoretical approximations.

\subsection{TR-Based Distance Estimator}
Since the shape of the TRRS distribution function $\bar{\eta}(d)\approx J_0^2(kd)$ is only determined by the wave number $k$ which is independent of specific locations, it can be utilized as an intrinsic ruler to measure distance in the space. Consider that an RX moves along a straight line with a constant speed $v$ starting from location $\vec{R}_0$, and a TX keeps transmitting the TR waveform corresponding to $\vec{R}_0$ at regular intervals. Then, the TRRS measured at the RX is just a sampled version of $\eta(d)$, which would also exhibit the Bessel-function-like pattern, as illustrated in Fig.~\ref{fig:Speed_Estimation_Algorithm}.

Take the first local peak of $\eta(d)$ for example. The corresponding theoretical distance $d_1$ is about $0.61\lambda$ according to the Bessel-function-like pattern. In order to estimate the moving speed, we only need to estimate how much time $\hat{t}$ it takes for the RX to reach the first local peak starting from point $\vec{R}_0$. We use a quadratic curve to approximate the shape of the first local peak. Combining the knowledge of the timestamps of each CIR measurement, $\hat{t}$ can be estimated by the vertex of the quadratic curve. Therefore, we obtain the speed estimation as $\hat{v}=(0.61\lambda)/\hat{t}$, and then, the moving distance can be calculated by integrating the instantaneous speed over time. One thing to note is that as long as the rate of CIR measurement is fast enough, it is reasonable to assume that the moving speed is constant during the measurement of the TRRS distribution. For example, in Fig.~\ref{fig:Speed_Estimation_Algorithm} the duration is about $0.16$ seconds.

Note that besides taking advantage of TR spatial focusing effect, the proposed distance estimator also exploits the physical properties of EM waves and thus does not require any pre-calibration, while the estimator presented in our previous work~\cite{Zhang17time} needs to measure the TRRS spatial decay curve in advance.


\section{Moving Direction Estimation and Error Correction}
\label{sec:Direction_Map}
In this section, we introduce the other two key components of WiBall: the IMU-based moving direction estimator and the map-based position corrector.

\subsection{IMU-based Moving Direction Estimator}
\label{subsec:IMU_Direction_Estimator}
\begin{figure}[!htbp]
	\centering
	\includegraphics[width=0.35\textwidth]{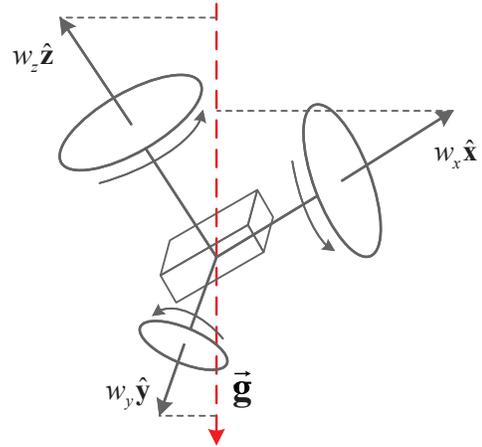}
	\caption{\label{fig:DirectionEstimation} Transforming the rotation of RX into the moving direction in horizontal plane.}
\end{figure}

If the RX is placed in parallel to the horizontal plane, the change of moving direction can be directly measured by the readings of the gyroscope in the $z$-axis, i.e., $\theta(t_i)=\omega_z(t_{i-1})(t_i-t_{i-1})$, where $\omega_z(t_{i-1})$ denotes the angular velocity of the RX with respect to $z$-axis in its local coordinate system at time slot $t_{i-1}$. However, in practice, the angle of the inclination between the RX and the horizontal plane is not zero, as shown in Fig.~\ref{fig:DirectionEstimation}, and WiBall needs to transform the rotation of the RX into the change of the moving direction in the horizontal plane. Since the direction of the gravity $\vec{\mathbf{g}}/\|\vec{\mathbf{g}}\|$ can be estimated by the linear accelerometer, the rotation of the RX in the horizontal plane, which is orthogonal to the $\vec{\mathbf{g}}/\|\vec{\mathbf{g}}\|$, can be obtained by projecting the angular velocity vector $\vec{\boldsymbol{\omega}} = \omega_x \hat{\mathbf{x}} + \omega_y \hat{\mathbf{y}} + \omega_z \hat{\mathbf{z}}$ with respect to its local coordinate system onto the direction $\vec{\mathbf{g}}/\|\vec{\mathbf{g}}\|$. Therefore, the change of moving direction $\theta(t_i)$ is obtained as
\begin{eqnarray}
\theta(t_i) = \frac{\vec{\boldsymbol{\omega}}^T(t_{i-1})\vec{\mathbf{g}}(t_{i-1})}{\|\vec{\mathbf{g}}(t_{i-1})\|} \cdot (t_i-t_{i-1}),
\end{eqnarray}
where $\vec{\boldsymbol{\omega}}(t_{i-1})$ and $\vec{\mathbf{g}}(t_{i-1})$ denote the vector of angular velocity and the gravity at time $t_{i-1}$, respectively.

\subsection{Map-based Position Corrector}
\label{subsec:Map_Based_Corrector}

\begin{figure}[t]
	\centering
	\includegraphics[width=0.4\textwidth]{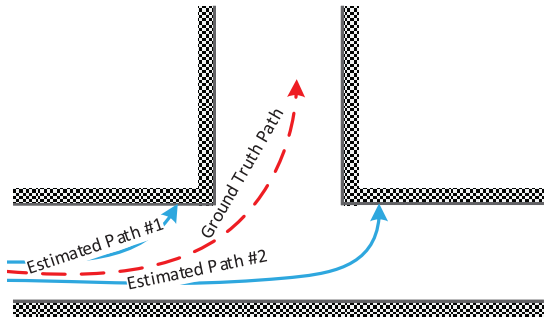}
	\caption{\label{fig:Map_Correction} Two possible estimated paths and the ground truth path.}
\end{figure}

Since WiBall estimates the current location of the RX based on the previous locations, its performance is limited by the cumulative error. However, for typical indoor environments, there are certain constraints in the floorplan which can be utilized as landmarks and thus, the cumulative errors may be corrected correspondingly as long as a landmark is identified. For example, Fig.~\ref{fig:Map_Correction} shows a T-shaped corridor and two possible estimated paths are illustrated in the figure. The moving distance of path \#1 is underestimated and that of path \#2 is overestimated, while the dotted line corresponds to the ground truth path. Both of the estimated traces would penetrate the wall in the floorplan if the errors are not corrected, which violates the physical constraints imposed by the structure of the building. In these two cases, a reasoning procedure can be implemented and WiBall tries to find the most possible path that can be fitted to the floorplan where all the border constraints imposed by the floorplan are satisfied. Therefore, the cumulative errors of both the distance estimations and direction estimations can be corrected when a map-based position correction is implemented.

\section{Performance Evaluation}
\label{sec:Experiment_Evaluations}
To evaluate the performance of WiBall, various experiments are conducted in different indoor environments using the prototype as shown in Fig.~\ref{fig:TR_prototype}. In this section, we first evaluate the performance of the TR-based distance estimator. Then, the performance of WiBall in tracking a moving object in two different environments is studied. At last, the impact of packet loss on the proposed system is also discussed.

\subsection{Evaluations of TR Distance Estimator}
\begin{figure}[htbp]
    \centering
    \begin{subfigure}[b]{0.3\textwidth}
        \includegraphics[width=\textwidth]{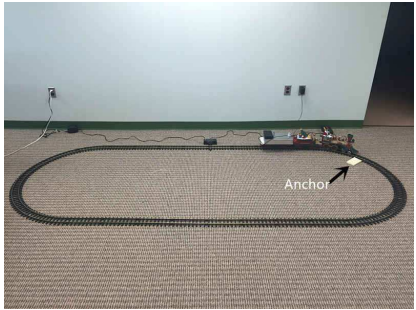}
        \caption{The toy train and the train track used in the experiment.}
        \label{fig:ToyTrack}
    \end{subfigure}
    ~
    \begin{subfigure}[b]{0.45\textwidth}
        \includegraphics[width=\textwidth]{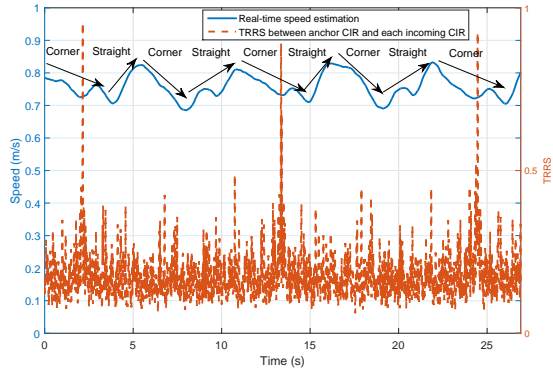}
	\caption{The estimated speed of the toy train over time.}
	\label{fig:Experiment_Toy_Train}
    \end{subfigure}
    \caption{Tracking the speed of the toy train.}
\end{figure}

\begin{figure}[t]
\centering
\includegraphics[width=0.45\textwidth]{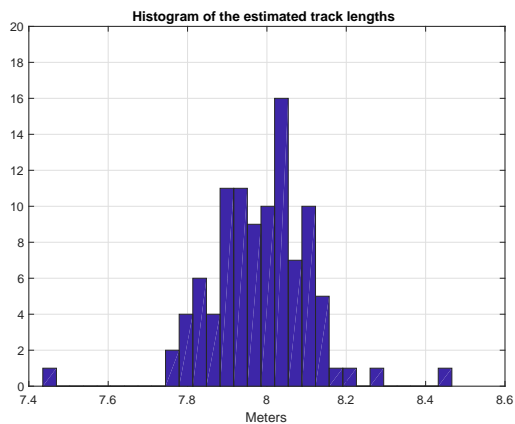}
\caption{The histogram of the estimated track lengths for a total of 100 experiments.}
\label{fig:HistToyTrain}
\end{figure}

The first experiment is to estimate the moving distance of a toy train running on a track. We put one RX on a toy train as shown in Fig.~\ref{fig:ToyTrack} and place one TX about $20$m from the RX with two walls between them. The sampling period between adjacent channel measurement is set to $T = 5\,$ms. CIRs are collected continuously when the toy train is running on the track. We also set an anchor point as shown in Fig.~\ref{fig:ToyTrack} on the train track and collect the CIR when the train is at the anchor. The TRRS values between all the measured CIRs and the CIR of the anchor are computed and shown in Fig.~\ref{fig:Experiment_Toy_Train}. The peaks in the red line indicate that the train passes the anchor three times. The estimated length of the track for this single loop is $8.12$m and the error is $1.50\%$, given the actual length of the train track is $8.00$m. The train slows down when it makes turns due to the increased friction and then speeds up in the straight line. This trend is reflected in the speed estimation shown by the blue curve. To show the consistency of the distance estimator over time, we collect the CIRs for 100 laps in total and estimate the track length for each lap separately. The histogram of the estimation results is shown in Fig.~\ref{fig:HistToyTrain}. The mean of the estimation error is about $0.02$m and the standard error deviation is about $0.13$m, which shows that the estimation is consistent even over a long period.

\begin{figure}[htbp]
    \centering
    \begin{subfigure}[b]{0.45\textwidth}
        \includegraphics[width=\textwidth]{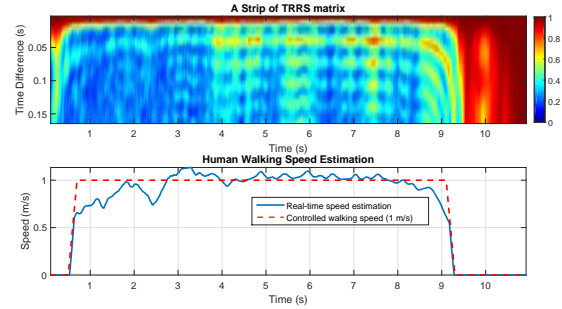}
        \caption{ Speed estimation for a controlled human walking speed ($1$m/s).}
        \label{fig:Experiment_Human_Walking}
    \end{subfigure}
    ~
    \begin{subfigure}[b]{0.45\textwidth}
        \includegraphics[width=\textwidth]{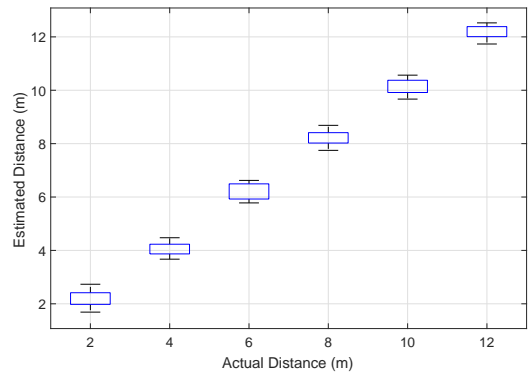}
\caption{The results for walking distance estimation.}
	\label{fig:ErrorBar}
    \end{subfigure}
    \caption{Human walking speed/distance estimation.}
\end{figure}

The second experiment is to estimate the human walking distance. One RX is put on a cart and one participant pushes the cart along the line from point A to point B shown in Fig.~\ref{fig:TR_prototype} with an approximately constant speed of $1$m/s. To control the walking speed, the participant uses a timer and landmarks placed on the floor during the experiment. In the upper panel of Fig.~\ref{fig:Experiment_Human_Walking}, for each time $t$, the TRRS values between the CIR measured at time $t$ and those measured at time $t-\Delta t$, where $\Delta t\in(0s,0.16s]$, are plotted along the vertical axis. As we can see from the figure, when the person moves slowly (e.g., at the beginning or the end of the experiment), the time differences between the local peaks of the measured TRRS distribution along the vertical axis are greater than that when the person moves faster. In addition, for $t\in[0.5s, 3.5s]$, the asymmetric part of the density function $P(\theta)$ of the energy of MPCs is more significant compared to the case when $t\in[3s,9.5s]$ and thus the pattern of $J^2(kd)$ is less obvious than the latter one. The bottom panel of Fig.~\ref{fig:Experiment_Human_Walking} shows the corresponding walking speed estimation. The actual distance is $8$m and the estimated walking distance is $7.65$m, so the corresponding error is $4.4\%$. The loss of performance is from the blockage of signals by the human body, which reduces the number of significant MPCs.

We further let the participant carry the RX and walk for a distance of $2$m, $4$m, $6$m, $8$m, $10$m and $12$m, respectively. For each ground-truth distance, the experiment is repeated $20$ times with different paths and the walking speed does not need to be constant. The results are shown in Fig.~\ref{fig:ErrorBar}, where the $5$, $25$, $75$, and $95$-th percentiles of the estimated distances for each actual distance are plotted from the bottom line to the top line for each block. We find that when the ground-truth distance is small, the error tends to be large. This is mainly because the participant could introduce additional sources of errors which are uncontrollable, such as not following the path strictly, shaking during walking, and not stopping at the exact point in the end. When the distance is short, the impact of this kind of error can be magnified greatly. However, when the walking distance is large, the impact of the uncontrollable errors on the estimation result is insignificant.

\subsection{Estimated Traces in Different Environment}
\label{subsec:Estimated_Traces_Different_Environment}

\begin{figure}[t]
    \centering
    \begin{subfigure}[b]{0.45\textwidth}
        \includegraphics[width=\textwidth]{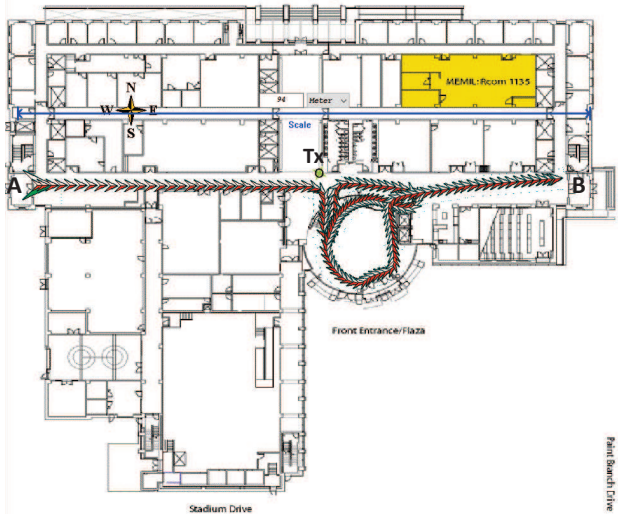}
        \caption{\label{fig:Kim_Experiment_Result} Estimated path in a building with a lot of open space.}
    \end{subfigure}
    ~
    \begin{subfigure}[b]{0.45\textwidth}
        \includegraphics[width=\textwidth]{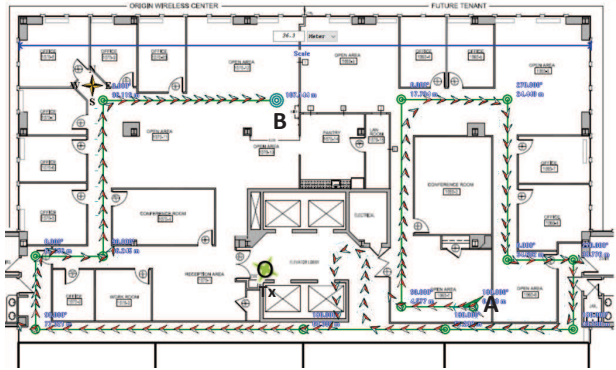}
		\caption{\label{fig:Office_Experiment_Result} Estimated path in an office environment.}
    \end{subfigure}
    \caption{\label{fig:Kim_Office_Experiment_Result} Experiment results in different environments.}
\end{figure}

We evaluate the performance of indoor tracking using WiBall in two sets of experiments. In the first set of experiments, a participant walks inside a building with a large open space. He carriers the RX with him and walks from Point A on the second floor to point B on the first floor, as shown in Fig.~\ref{fig:Kim_Experiment_Result}. The TX is placed closed to the middle of the path on the second floor. The dimension of the building is around $94\,m\times 73\,m$. Although the moving distance of the first segment of the path is overestimated, the estimated path is corrected when the participant enters the staircase leading to the first floor.

In the second set of experiments, the participant walks inside an office environment. Fig.~\ref{fig:Office_Experiment_Result} demonstrates a typical example of the estimated traces in a typical office of a multi-storey. One RX is put on a cart and the participant pushes the cart along the route from Point A to Point B, as illustrated in the figure. The dimension of the environment is around $36.3\,m\times 19\,m$ and the placement of the TX is also shown in the figure. As we can see from the figure, the estimated path matches the ground truth path very well because the cumulative errors have been corrected by the constraints from the floorplan.

\subsection{Statistical Analysis of Localization Error}
\label{subsec:Statistical_Analysis_Localization_Error}
\begin{figure}[t]
	\centering
	\includegraphics[width=0.4\textwidth]{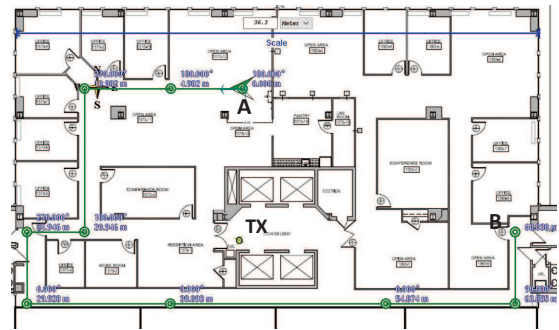}
	\caption{\label{fig:Statistical_Error} The route for the evaluation of statistical errors.}
\end{figure}

To evaluate the distribution of the localization errors, extensive experiments have been conducted in the same office environment shown in Fig.~\ref{fig:Office_Experiment_Result}. The participant pushes the cart with the RX on the cart, following the route as shown in Fig.~\ref{fig:Statistical_Error}.

\begin{figure}[t]
	\centering
	\includegraphics[width=0.4\textwidth]{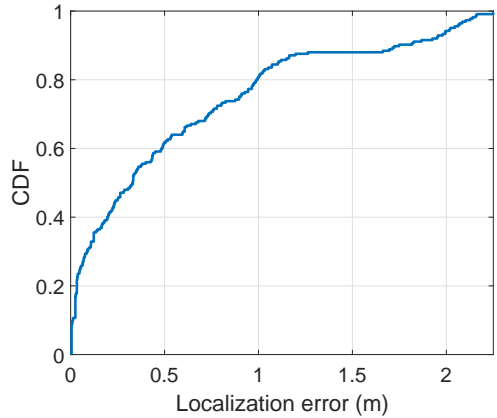}
	\caption{\label{fig:CDF_Localization_Error} Empirical CDF of localization error.}
\end{figure}

The RX starts from Point A and stops at different locations in the route shown in Fig.~\ref{fig:Statistical_Error}. The lengths of the chosen paths are $5$, $21$, $25$, $30$, $40$, $64$, and $69$m, respectively, and the end of each path is marked with two green circles. For each specific path, the experiment is repeated for $25$ times. The estimation error for different paths has been analyzed through empirical cumulative distribution function (CDF), as shown in Fig.~\ref{fig:CDF_Localization_Error}. Based on the results, the median of the estimation error for the selected paths is around $0.33$m, and the $80$ percentile of the estimation error is around $1$m. Therefore, WiBall is able to achieve a sub-meter median error in this complex indoor environment.

\subsection{Impact of Packet Loss on Distance Estimation}
In the previous experiments, WiBall operates on a vacant band and the packet loss rate can thus be neglected. However, in practice, the RF interference from other RF devices operated on the same frequency band will increase the packet loss rate. Since WiBall relies on the first peak of the TRRS distribution for distance estimation, enough samples need to be collected so as to estimate the first peak accurately, and a high packet loss can affect the peak estimation and thus increase distance estimation error.

\begin{figure}[htbp]
    \centering
    \begin{subfigure}[b]{0.23\textwidth}
        \includegraphics[width=\textwidth]{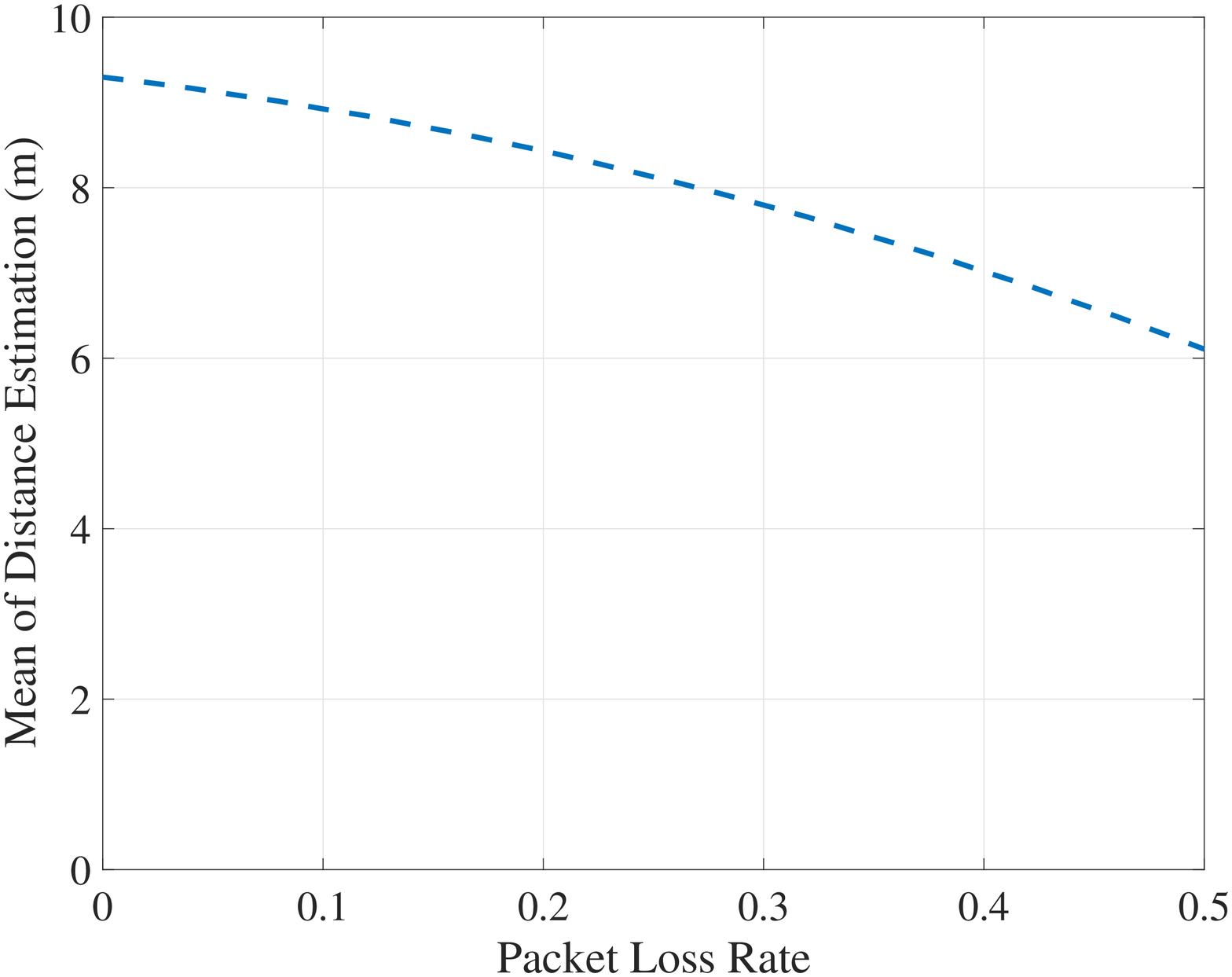}
        \caption{\label{fig:Mean_Estimation_Packet_Loss} Sample mean of the distance estimation.}
    \end{subfigure}
    ~
    \begin{subfigure}[b]{0.23\textwidth}
        \includegraphics[width=\textwidth]{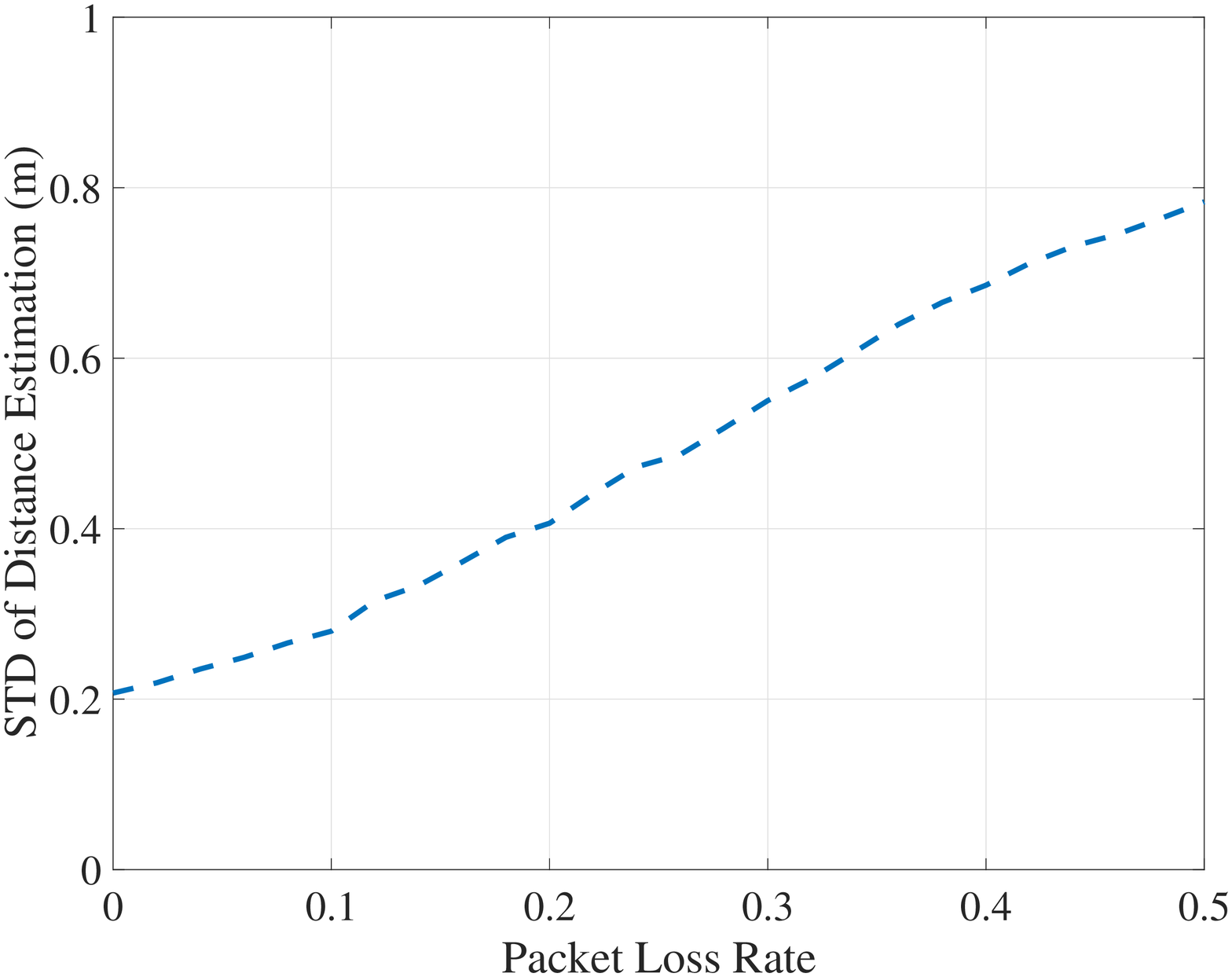}
		\caption{\label{fig:STD_Estimation_Packet_Loss} Standard deviation of the distance estimation.}
    \end{subfigure}
    \caption{\label{fig:Packet_Loss} The impact of packet loss on the accuracy of distance estimation.}
\end{figure}

To study the impact of RF interference, a pair of RF devices is configured to operate in the same frequency band of WiBall to act as an interference source, and we run the tracking system for $100$ times with the ground-truth distance being $10$m. When the interfering devices are placed closer to the transmission pair of WiBall, WiBall encounters a higher packet loss rate. Therefore, to obtain various packet loss rates, the interfering devices are placed at different locations during the experiment. The average estimated distance and standard deviation of the estimation under different packet loss rates are shown in Fig.~\ref{fig:Packet_Loss}. It is seen that a large packet loss rate would lead to an underestimation of the moving distance and increase the deviations of the estimates.

\section{Conclusions}
\label{sec:Concluding_Remarks}
In this work, we propose WiBall, which offers an accurate, low-cost, calibration-free and robust solution for INIP in indoor environments without requiring any infrastructure. WiBall leverages the physical properties of the TR focusing ball in radio signals, such as WiFi, LTE, 5G, etc., to estimate the moving distance of an object being monitored and does not need any specialized hardware. It is shown through extensive experiments that WiBall can achieve a decimeter-level accuracy. WiBall is also easily scalable and can accommodate a large number of users with only a single access point or TX. Therefore, WiBall can be a very promising candidate for future indoor tracking systems.

\bibliographystyle{ieeetr}
\bibliography{main}

\begin{thebibliography}{10}

\bibitem{report17market}
``{Global Indoor Positioning and Indoor Navigation (IPIN) Market 2017-2021}.''
  \url{https://www.technavio.com/report/global-machine-machine-m2m-and-connected-devices-global-indoor-positioning-and-indoor?utm_source=T3&utm_medium=BW&utm_campaign=Media.
  [Accessed: 07-Dec-2017}, 2017.
\newblock [Online].

\bibitem{chen2017achieving}
C.~Chen, Y.~Chen, Y.~Han, H.-Q. Lai, and K.~J.~R. Liu, ``Achieving
  centimeter-accuracy indoor localization on wifi platforms: A frequency
  hopping approach,'' {\em IEEE Internet of Things Journal}, vol.~4, no.~1,
  pp.~111--121, 2017.

\bibitem{werner2011indoor}
M.~Werner, M.~Kessel, and C.~Marouane, ``Indoor positioning using smartphone
  camera,'' in {\em Indoor Positioning and Indoor Navigation (IPIN), 2011
  International Conference on}, pp.~1--6, IEEE, 2011.

\bibitem{mautz2011survey}
R.~Mautz and S.~Tilch, ``Survey of optical indoor positioning systems,'' in
  {\em Indoor Positioning and Indoor Navigation (IPIN), 2011 International
  Conference on}, pp.~1--7, IEEE, 2011.

\bibitem{gorostiza2011infrared}
E.~M. Gorostiza, J.~L. L{\'a}zaro~Galilea, F.~J. Meca~Meca,
  D.~Salido~Monz{\'u}, F.~Espinosa~Zapata, and L.~Pallar{\'e}s~Puerto,
  ``Infrared sensor system for mobile-robot positioning in intelligent
  spaces,'' {\em Sensors}, vol.~11, no.~5, pp.~5416--5438, 2011.

\bibitem{rishabh2012indoor}
I.~Rishabh, D.~Kimber, and J.~Adcock, ``Indoor localization using controlled
  ambient sounds,'' in {\em Indoor Positioning and Indoor Navigation (IPIN),
  2012 International Conference on}, pp.~1--10, IEEE, 2012.

\bibitem{barton2004radar}
D.~K. Barton, {\em Radar system analysis and modeling}, vol.~1.
\newblock Artech House, 2004.

\bibitem{kuhn2008high}
M.~Kuhn, C.~Zhang, B.~Merkl, D.~Yang, Y.~Wang, M.~Mahfouz, and A.~Fathy, ``High
  accuracy uwb localization in dense indoor environments,'' in {\em
  Ultra-Wideband, 2008. ICUWB 2008. IEEE International Conference on}, vol.~2,
  pp.~129--132, IEEE, 2008.

\bibitem{lee2002ranging}
J.-Y. Lee and R.~A. Scholtz, ``Ranging in a dense multipath environment using
  an uwb radio link,'' {\em IEEE Journal on Selected Areas in Communications},
  vol.~20, no.~9, pp.~1677--1683, 2002.

\bibitem{wang2012no}
H.~Wang, S.~Sen, A.~Elgohary, M.~Farid, M.~Youssef, and R.~R. Choudhury, ``No
  need to war-drive: Unsupervised indoor localization,'' in {\em Proceedings of
  the 10th international conference on Mobile systems, applications, and
  services}, pp.~197--210, ACM, 2012.

\bibitem{yang2015mobility}
Z.~Yang, C.~Wu, Z.~Zhou, X.~Zhang, X.~Wang, and Y.~Liu, ``Mobility increases
  localizability: A survey on wireless indoor localization using inertial
  sensors,'' {\em ACM Computing Surveys (Csur)}, vol.~47, no.~3, p.~54, 2015.

\bibitem{giustiniano2011caesar}
D.~Giustiniano and S.~Mangold, ``Caesar: carrier sense-based ranging in
  off-the-shelf 802.11 wireless lan,'' in {\em Proceedings of the Seventh
  COnference on emerging Networking EXperiments and Technologies}, p.~10, ACM,
  2011.

\bibitem{sen2015bringing}
S.~Sen, D.~Kim, S.~Laroche, K.-H. Kim, and J.~Lee, ``Bringing cupid indoor
  positioning system to practice,'' in {\em Proceedings of the 24th
  International Conference on World Wide Web}, pp.~938--948, International
  World Wide Web Conferences Steering Committee, 2015.

\bibitem{bahl2000radar}
P.~Bahl and V.~N. Padmanabhan, ``Radar: An in-building rf-based user location
  and tracking system,'' in {\em INFOCOM 2000. Nineteenth Annual Joint
  Conference of the IEEE Computer and Communications Societies. Proceedings.
  IEEE}, vol.~2, pp.~775--784, Ieee, 2000.

\bibitem{kotaru2015spotfi}
M.~Kotaru, K.~Joshi, D.~Bharadia, and S.~Katti, ``Spotfi: Decimeter level
  localization using wifi,'' in {\em ACM SIGCOMM Computer Communication
  Review}, vol.~45, pp.~269--282, ACM, 2015.

\bibitem{gjengset2014phaser}
J.~Gjengset, J.~Xiong, G.~McPhillips, and K.~Jamieson, ``Phaser: Enabling
  phased array signal processing on commodity wifi access points,'' in {\em
  Proceedings of the 20th annual international conference on Mobile computing
  and networking}, pp.~153--164, ACM, 2014.

\bibitem{sen2013avoiding}
S.~Sen, J.~Lee, K.-H. Kim, and P.~Congdon, ``Avoiding multipath to revive
  inbuilding wifi localization,'' in {\em Proceeding of the 11th annual
  international conference on Mobile systems, applications, and services},
  pp.~249--262, ACM, 2013.

\bibitem{youssef2005horus}
M.~Youssef and A.~Agrawala, ``The horus wlan location determination system,''
  in {\em Proceedings of the 3rd international conference on Mobile systems,
  applications, and services}, pp.~205--218, ACM, 2005.

\bibitem{xiong2015tonetrack}
J.~Xiong, K.~Sundaresan, and K.~Jamieson, ``Tonetrack: Leveraging
  frequency-agile radios for time-based indoor wireless localization,'' in {\em
  Proceedings of the 21st Annual International Conference on Mobile Computing
  and Networking}, pp.~537--549, ACM, 2015.

\bibitem{golden2007sensor}
S.~A. Golden and S.~S. Bateman, ``Sensor measurements for wi-fi location with
  emphasis on time-of-arrival ranging,'' {\em IEEE Transactions on Mobile
  Computing}, vol.~6, no.~10, 2007.

\bibitem{castro2001probabilistic}
P.~Castro, P.~Chiu, T.~Kremenek, and R.~Muntz, ``A probabilistic room location
  service for wireless networked environments,'' in {\em Ubicomp 2001:
  Ubiquitous Computing}, pp.~18--34, Springer, 2001.

\bibitem{wang2015deepfi}
X.~Wang, L.~Gao, S.~Mao, and S.~Pandey, ``Deepfi: Deep learning for indoor
  fingerprinting using channel state information,'' in {\em Wireless
  Communications and Networking Conference (WCNC), 2015 IEEE}, pp.~1666--1671,
  IEEE, 2015.

\bibitem{wang2015phasefi}
X.~Wang, L.~Gao, and S.~Mao, ``Phasefi: Phase fingerprinting for indoor
  localization with a deep learning approach,'' in {\em Global Communications
  Conference (GLOBECOM), 2015 IEEE}, pp.~1--6, IEEE, 2015.

\bibitem{wu2015time}
Z.-H. Wu, Y.~Han, Y.~Chen, and K.~J.~R. Liu, ``A time-reversal paradigm for
  indoor positioning system,'' {\em IEEE Transactions on Vehicular Technology},
  vol.~64, no.~4, pp.~1331--1339, 2015.

\bibitem{xiong2013arraytrack}
J.~Xiong and K.~Jamieson, ``Arraytrack: A fine-grained indoor location
  system,'' Usenix, 2013.

\bibitem{wu2013csi}
K.~Wu, J.~Xiao, Y.~Yi, D.~Chen, X.~Luo, and L.~M. Ni, ``Csi-based indoor
  localization,'' {\em IEEE Transactions on Parallel and Distributed Systems},
  vol.~24, no.~7, pp.~1300--1309, 2013.

\bibitem{sen2011precise}
S.~Sen, B.~Radunovic, R.~Roy~Choudhury, and T.~Minka, ``Precise indoor
  localization using phy information,'' in {\em Proceedings of the 9th
  international conference on Mobile systems, applications, and services},
  pp.~413--414, ACM, 2011.

\bibitem{park2003level}
G.~Park, D.~Hong, and C.~Kang, ``Level crossing rate estimation with doppler
  adaptive noise suppression technique in frequency domain,'' in {\em Vehicular
  Technology Conference, 2003. VTC 2003-Fall. 2003 IEEE 58th}, vol.~2,
  pp.~1192--1195, IEEE, 2003.

\bibitem{sampath1993estimation}
A.~Sampath and J.~M. Holtzman, ``Estimation of maximum doppler frequency for
  handoff decisions,'' in {\em Vehicular Technology Conference, 1993., 43rd
  IEEE}, pp.~859--862, IEEE, 1993.

\bibitem{xiao2001mobile}
C.~Xiao, K.~D. Mann, and J.~C. Olivier, ``Mobile speed estimation for
  tdma-based hierarchical cellular systems,'' {\em IEEE Transactions on
  Vehicular Technology}, vol.~50, no.~4, pp.~981--991, 2001.

\bibitem{narasimhan1999speed}
R.~Narasimhan and D.~C. Cox, ``Speed estimation in wireless systems using
  wavelets,'' {\em IEEE Transactions on Communications}, vol.~47, no.~9,
  pp.~1357--1364, 1999.

\bibitem{lerosey2004time}
G.~Lerosey, J.~De~Rosny, A.~Tourin, A.~Derode, G.~Montaldo, and M.~Fink, ``Time
  reversal of electromagnetic waves,'' {\em Physical review letters}, vol.~92,
  no.~19, p.~193904, 2004.

\bibitem{wang2011green}
B.~Wang, Y.~Wu, F.~Han, Y.-H. Yang, and K.~J.~R. Liu, ``Green wireless
  communications: A time-reversal paradigm,'' {\em IEEE Journal on Selected
  Areas in Communications}, vol.~29, no.~8, pp.~1698--1710, 2011.

\bibitem{roux1997time}
P.~Roux, B.~Roman, and M.~Fink, ``Time-reversal in an ultrasonic waveguide,''
  {\em Applied Physics Letters}, vol.~70, no.~14, pp.~1811--1813, 1997.

\bibitem{el2010experimental}
H.~El-Sallabi, P.~Kyritsi, A.~Paulraj, and G.~Papanicolaou, ``Experimental
  investigation on time reversal precoding for space--time focusing in wireless
  communications,'' {\em IEEE Transactions on Instrumentation and Measurement},
  vol.~59, no.~6, pp.~1537--1543, 2010.

\bibitem{Zhang17time}
F.~Zhang, C.~Chen, B.~Wang, H.-Q. Lai, and K.~J.~R. Liu, ``A time-reversal
  spatial hardening effect for indoor speed estimation,'' in {\em Proc. of IEEE
  ICASSP}, pp.~5955--5959, March 2017.

\end{thebibliography}

\end{document}